\documentclass[%
 prx,
 amsmath,amssymb,
 reprint,%
 showkeys,
 floatfix
]{revtex4-2}

\usepackage{graphicx}% Include figure files
\usepackage{dcolumn}% Align table columns on decimal point
\usepackage{bm}% bold math
\usepackage{nicefrac}
\usepackage[mathlines]{lineno}% Enable numbering of text and display math
\usepackage{xr-hyper} %cross reference to figures in supplementary information
\usepackage{hyperref}
\hypersetup{%
    colorlinks = true,
    linkcolor = black,
    urlcolor = black,
    citecolor = black,
}
\usepackage[capitalise,nameinlink]{cleveref}
\usepackage[utf8]{inputenc}
\usepackage[T1]{fontenc}
\usepackage{mathptmx}
\usepackage{etoolbox}
\usepackage[version=3]{mhchem} % Formula subscripts using \ce{}
\usepackage{cancel}
\usepackage[table]{xcolor}
\usepackage{mathtools}
\usepackage{booktabs}
\usepackage{csquotes}
\newcommand*\linenomathpatch[1]{%
  \cspreto{#1}{\linenomath}%
  \cspreto{#1*}{\linenomath}%
  \csappto{end#1}{\endlinenomath}%
  \csappto{end#1*}{\endlinenomath}%
}

\linenomathpatch{equation}
\linenomathpatch{gather}
\linenomathpatch{multline}
\linenomathpatch{align}
\linenomathpatch{alignat}
\linenomathpatch{flalign}
\definecolor{LightCyan}{rgb}{0.88,1,1}
\definecolor{lightgray}{gray}{0.9}

\newcommand{\iu}{{i\mkern1mu}}

\newcommand{\bk}{{\mathbf{k}}}

\newcommand{\bq}{{\mathbf{q}}}
\newcommand{\bQ}{{\mathbf{Q}}}
\newcommand{\br}{{\mathbf{r}}}
\newcommand{\bR}{{\mathbf{R}}}

\newcommand{\bu}{{\mathbf{u}}}

\newcommand{\qty}[2]{{$#1\mskip3mu$\text{#2}}}

\DeclareMathAlphabet{\mymathbb}{U}{BOONDOX-ds}{m}{n}
\makeatletter
\def\@email#1#2{%
 \endgroup
 \patchcmd{\titleblock@produce}
  {\frontmatter@RRAPformat}
  {\frontmatter@RRAPformat{\produce@RRAP{*#1\href{mailto:#2}{#2}}}\frontmatter@RRAPformat}
  {}{}
}%
\newcommand*{\addFileDependency}[1]{% argument=file name and extension
  \typeout{(#1)}
  \@addtofilelist{#1}
  \IfFileExists{#1}{}{\typeout{No file #1.}}
}
\makeatother

\newcommand*{\myexternaldocument}[1]{%
    \externaldocument{#1}%
    \addFileDependency{#1.tex}%
    \addFileDependency{#1.aux}%
}
\myexternaldocument{polaron_SI}
\begin{document}

\preprint{APS/123-QED}

\title[A momentum-resolved view of polaron formation in materials]{A momentum-resolved view of polaron formation in materials}% Force line breaks with \\

\author{Tristan L Britt}
\email{tristan.britt@mail.mcgill.ca}
\affiliation{Department of Physics, Center for the Physics of Materials, McGill University, 3600 rue Université, Montréal, Québec H3A 2T8, Canada}

\author{Fabio Caruso}
\email{caruso@physik.uni-kiel.de}
\affiliation{Institut für Theoretische Physik und Astrophysik, Christian-Albrechts-Universität zu Kiel, D-24118 Kiel, Germany}
\affiliation{Kiel Nano, Surface and Interface Science KiNSIS, Christian-Albrechts-Universität zu Kiel, D-24118 Kiel, Germany}

\author{Bradley J Siwick}%
\email{bradley.siwick@mcgill.ca}
\affiliation{%
 Department of Physics, Center for the Physics of Materials, McGill University, 3600 rue Université, Montréal, Québec H3A 2T8, Canada}
\affiliation{Department of Chemistry, McGill University, 801 rue Sherbrooke Ouest, Montréal, Québec H3A 0B8, Canada}

\date{\today}% It is always \today, today,

\begin{abstract}
A combined experimental and computational methodology for interrogating the phonon contribution to polaron formation in real materials is developed. Using LiF as an example, we show that the recent ab-initio theory of Sio et. al [PRL \textbf{122}, 246403 (2019)] makes predictions of the momentum- and branch dependent phonon amplitides in polaron quasiparticles that are testable using ultrafast electron diffuse scattering (UEDS) and related techniques. The large electron polaron in LiF has UEDS signatures that are qualitatively similar to those expected from a simple point-defect model, but the small hole polaron exhibits a profoundly anisotropic UEDS pattern that is in poor agreement with a point-defect model. We also show that these polaron diffuse scattering signatures are directly emblematic of the underlying polaron wavefunction.  The combination of new time and momentum resolved experimental probes of nonequilibrium phonons with novel computational methods promises to complement the qualitative results obtained via model Hamiltonians with a first principles, material-specific quantiative understanding of polarons and their properties.

\end{abstract}

\keywords{polarons, phonons, scattering, ultrafast diffraction, density functional theory}%Use showkeys class option if keyword
\maketitle

%\tableofcontents
\section{Introduction}

Developing an understanding of the quantum nature of an excess charge carrier
interacting with a polarizable environment is a fundamental many-body problem
in condensed matter physics \cite{Franchini2021,Ren2023} and chemistry
\cite{RevModPhys.65.599}.  In polar semiconducting and insulating materials,
excess electrons and holes can couple to lattice vibrations in such a way that
yields a quasiparticle called a polaron;  i.e. an electron/hole that is
screened (or dressed) by phonons, localizing the charge and reducing its energy
\cite{Franchini2021}.  Transport of these quasiparticles involves both the
localized carrier and its associated lattice distortion (phonon-cloud).  Such
phonon-polarons are a long-standing example of quasiparticles formed through
the coupling of an excess carrier to a boson-bath, which are fundamental to a
range of processes and properties in materials including charge transport
\cite{Coropceanu2007,Natanzon2020,Nelson2009,Ortman2011}, superconductivity
\cite{Salje1995}, light-harvesting in novel perovskite materials
\cite{miyata2017large, ghosh2020polarons, guzelturk2021visualization} and
thermoelectricity \cite{Wang2014}.  More recently, the polaron concept has 
been broadened through the discovery of Fermi-polarons in 2D materials,  which
involve coupling of carriers to a fermi-bath of electron/hole pair excitations
rather than phonons \cite{koschorreck2012attractive, muir2022interactions}, 
and plasmonic polarons \cite{caruso_band_2015}, 
which arise from coupling to carrier plasmons in highly-doped 
semiconductors and insulators 
\cite{riley_crossover_2018,caruso_two-dimensional_2021,ma_formation_2021,emeis_plasmonic_2023}. 

The polaron problem has provided strong impulses for advancing the 
fundamental understanding of many-body interactions and electron-phonon 
interactions in condensed-matter systems \cite{Franchini2021}. 
Much of the early understanding of phonon-polarons has been developed
through the study of simplified models due to Fr\"ohlich \cite{Frolich1954} and
Holstein \cite{Holstein1959} that provide a description of free or tightly
bound electrons coupled to a single dispersionless optical phonon, and have
helped to define the limits of large and small polarons respectively.
The development of \emph{ab-initio} approaches for studying polarons in realistic atomistic models of materials
has initially relied on Kohn-Sham density-functional theory \cite{Kohn_PhysRev.140.A1133}, whereby polarons 
can be described by explicitly modelling excess charges in a 
supercell model \cite{janotti_PhysRevB.90.085202,Franchini_PhysRevLett.102.256402} with suitable self-interaction corrections
\cite{Pasquarello_PhysRevLett.129.126401,Pasquarello_PhysRevB.106.125119}. 
Subsequently, the formulation of an efficient scheme to handle long-range (Fr\"ohlich)
electron-phonon interactions \cite{Verdi2015} facilitated the extension of 
perturbative many-body approaches \cite{Verdi2015,verdi_origin_2017,Gonze_PhysRevB.97.115145}
to the polaron problem. 
Lately, the development of self-consistent density-functional
\cite{Sio2019_PRL,Sio2019,gonze_PhysRevB.104.235123} and many-body approaches
\cite{lafuente-bartolome_ab_2022,lafuente-bartolome_unified_2022} has further 
extended the theoretical understanding of polarons and the computational 
capabilities for their description, enabling the capture of multiple facets of 
polaron formation (e.g., electron localization, lattice distortion, 
and formation energies) without resorting to supercells or perturbative approximations. 

In parallel with these theoretical developments, new ultrafast experimental
probes of the electron and phonon systems in materials have become available;
e.g.  time and angle resolved photoelectron spectroscopy (trARPES)
\cite{Boschini2024}, ultrafast electron diffuse scattering (UEDS) \cite{RenedeCotret2022,Helene2021,Chase2016,Britt2022,Waldecker2017} and
its xray analog \cite{trigo2013fourier,Hartley2021}.  These methods provide time and momentum-resolved
information on the occupancy of electronic states and phonon modes in
pump-probe style experiments, opening up the possibility of studying polaron
formation at a truly unprecedented level of detail. Static ARPES has already
provided important information on the spectral fingerprints of polarons in anatase 
TiO$_2$ \cite{moser2013tunable}, SrTiO$_3$ \cite{wang_tailoring_2016}, 
and EuO \cite{riley_crossover_2018}, revealing the emergence of polaronic 
in-gap states in these materials. These earlier studies contributed to the identification of the spectral fingerprints that are expected from polaron formation in photoemission experiments, which allow for the
unambiguous detection and characterization of the \textit{electronic} properties of polaronic
quasiparticles in materials.  The \textit{lattice} properties of a polarons, however,
are not directly accesible from photoemission.  Diffuse scattering techniques
are inherently sensitive to lattice structure and its fluctuations, and in principle they are suitable to
directly probe polaronic distortions.  However, to date a first-principles
theory of polaron effects in diffuse scattering is still missing.  Thus,
it is in general not known which diffuse scattering signatures are unequivocal evidence of polarons nor how such signals might be expected to vary from material to material. While phenomenological models have been proposed in the past, these approaches are not predictive or quantitative descriptions of polaron formation \cite{Shimomura1999,guzelturk2021visualization}.

In this work, we formulate an \emph{ab-initio} approach to predict the 
fingerprints of polarons in diffuse scattering experiments. 
Specifically, we show that ultrafast electron diffuse scattering measurements
are directly sensitive to the polaron envelop function, and they thus enable
the characterization of polaron-induced structural changes in reciprocal space. 
To illustrate the predictive capability of this approach, we present
\emph{ab-initio} calculation of polaron diffuse scattering in LiF.
Our numerical simulations 
combine the solution of the self-consistent polaron equations -- 
as defined in \cite{Sio2019_PRL,Sio2019} -- with \emph{ab-initio} calculations 
of the dynamical structure factor \cite{Zacharias2021_PRB,Zacharias_PhysRevLett.127.207401}. 
Together, these computational and experimental
tools can provide quantitative insights into 
the nature and properties of polarons in real materials.  

The structure of the paper is as follows. In Sec.~\ref{sec:theo}, we summarize the 
self-consistent polaron equations \cite{Sio2019_PRL,Sio2019}, 
and introduce the phonon-diffuse scattering that are relevant to diffuse
scattering from polarons observable using UEDS. We make a quantitative
connection between the UEDS signals and the theoretical phonon amplitudes that
are core to this theory, suggesting the ease with which experiment and theory
can be compared.  In Sec.~\ref{sec:LiF}, we describe a framework for computing phonon diffuse
scattering patterns from the computed polarons, with LiF as the example. 
Our final remarks are reported in Sec.~\ref{sec:con}.

\section{Theoretical background}\label{sec:theo}
In this section, we define an \emph{ab-initio} computational procedure to predicatively infer the 
signatures of polarons in diffuse scattering experiments. Our approach consists of 
two steps: we first evaluate polaronic structural changes based on the 
\emph{ab-initio} polaron theory, as formulated by Sio \emph{et al} \cite{Sio2019}; 
subsequently, we integrate polaronic structural changes into the evaluation of the 
dynamical structure factor. 

\subsection{\emph{Ab-initio} polaron theory}
Below, we briefly review the key elements of the \emph{ab-initio} polaron theory 
relevant for the description of polarons in diffuse scattering. 
In the independent particle picture, the wave function 
of an excess electron (or hole) $\psi$ can be expressed 
as a linear superposition of Kohn-Sham orbitals $\psi_{n{\bf k}}$: 
\begin{align}\label{eq:pol1}
\psi=N_p^{-1 / 2} \sum_{n \mathbf{k}} A_{n \mathbf{k}} \psi_{n \mathbf{k}} \quad, 
\end{align}
where $A_{n\bk}$ is a complex envelop function, the sum runs over all 
band indices and crystal momenta in the Brillouin zone, and $N_p$ is the number of 
unit cells in the Born-von-Kármán supercell. 
In polar crystals, the interaction between the excess charge and the lattice can result 
into charge localization in concomitance with symmetry breaking structural changes (polaronic 
distortions). The atomic displacement from equilibrium can be expressed as: 
\begin{align} \label{eq:pol2}
\Delta {\boldsymbol \tau}_{p\kappa}^{\rm pol}=-2 N_p^{-1} \sum_{\mathbf{q} \nu} B_{\mathbf{q} \nu}^*\sqrt{\frac{\hbar}{2 M_\kappa \omega_{\mathbf{q} \nu}}} {\bf e}_{ \mathbf{q} \nu\kappa} e^{i \mathbf{q} \cdot \mathbf{R}_p}\quad. 
\end{align}
The indices $\kappa$  and $p$ label the atoms and unit cells, respectively.  $
B_{\mathbf{q} \nu}$ is a complex envelop function,  $M_\kappa$ is the atomic
mass, $\omega_{\mathbf{q} \nu}$ the phonon frequency, ${\bf e}_{
\mathbf{q} \nu\kappa}$ the phonon eigenvector, and  $\mathbf{R}_p$ a crystal lattice
vector. 

With the exception of $A_{n \mathbf{k}}$ and $B_{\mathbf{q} \nu}$, 
all quantities entering Eqs.~\eqref{eq:pol1} and \eqref{eq:pol2} can be 
obtained from first-principles calculations based on density-functional theory (DFT)
and  density-functional perturbation theory (DFPT). 
A possible route to study the formation of polarons in materials consists in finding the values for the 
envelop functions $A_{n \mathbf{k}}$ and $B_{\mathbf{q} \nu}$ that 
minimize the total energy of the system in presence of an excess charge. 
This variational problem can be reformulated as 
a self-consistent eigenvalue problem for $A_{n \mathbf{k}}$ and $B_{\mathbf{q} \nu}$ 
that must be solved iteratively: 
%These weights are shown to satisfy the self-consistent eigenvalue problem:
\begin{subequations}
  \label{eqn:polaron}
  \begin{equation}
    \frac{2}{N_p}\sum_{\bq m \nu}B_{\bq\nu}\left(g_{mn}^\nu(\bk, \bq)\right)^*A_{m\bk+\bq} = (\varepsilon_{n\bk}-\varepsilon)A_{n\bk}
  \end{equation}
  \begin{equation}
    B_{\bq\nu} = \frac{1}{N_p}\sum_{mn\bk}A^*_{m\bk+\bq}\frac{g_{mn}^\nu(\bk,\bq)}{\hbar\omega_{\bq\nu}}A_{n\bk}
  \end{equation}  
\end{subequations}
where $\varepsilon$ is the polaron eigenvalue, 
and $g_{mn}^\nu(\bk, \bq)$ is the electron-phonon coupling matrix element. 

In practice, Eq.~\eqref{eqn:polaron}  are solved self-consistently by
instantiating the electronic weights with Gaussian distributions
in the vicinity of valence band maximum (conduction band minimum) 
for the hole (electron) polaron. Further details can be found elsewhere \cite{Sio2019}. 
In this work, we solve the self-consistent polaron equations 
for LiF in order to quantitatively assess the emergence of 
polaronic effects. A detailed description of the computational settings is
reported in the \hyperref[sec:methods]{Methods}.

\begin{figure*}[!t]
  \includegraphics[width=\linewidth]{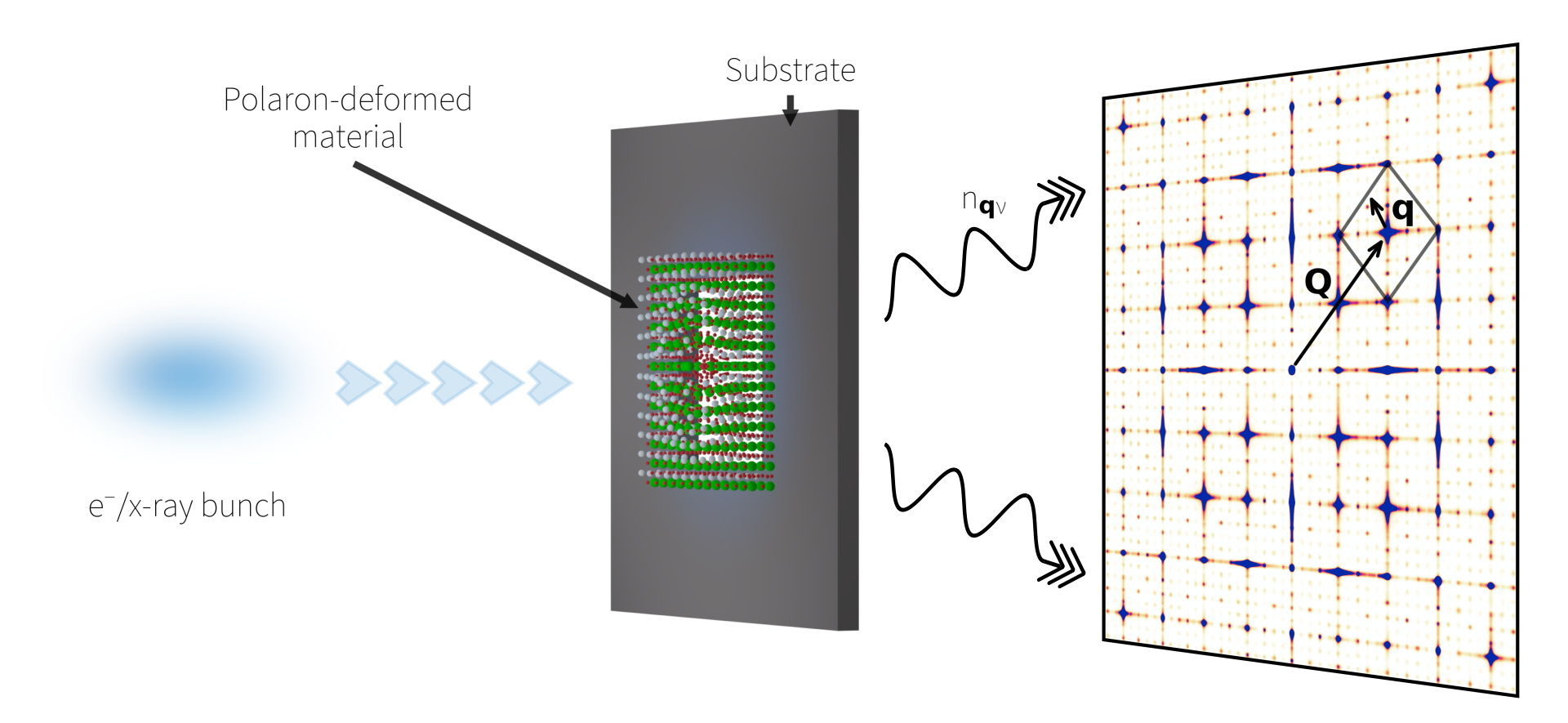}
  \caption{Representative schematic of ultrafast diffraction experiments on a polaronic material. The example given is cubic \ce{SrTiO3} with the ionic distortions of the electron polaron \cite{Natanzon2020}. The magnitudes of the displacements are exaggerated for illustration. The ionic distortions of the polaron are composed of linear combinations of phonons, with phonon momentum and mode dependence, the occupation of which is the experimental observable of ultrafast diffuse scattering. The diffuse scattering pattern (in the kinematic approximation) of equilibrium cubic \ce{SrTiO3} is given on the right, with the scattering vector (momentum transfer) $\bQ$ and a representative Brilluoin zone (BZ) illustrated. The location in the BZ indicative of phonon occupation at momentum $\bq$ is shown.}
  \label{fig:schematic}
\end{figure*}
\subsection{Phonon Diffuse Scattering}

We proceed to discuss the inclusion of   
structural distortions arising from the polaron formation within \emph{ab-initio} calculations of the inelastic scattering intensity.
The intensity of 
a wave scattered by the atoms in a crystalline lattice 
can be expressed within the Laval-Born-James theory \cite{Laval1939,Born1942,James1948} as: 
\begin{align}   \label{eqn:I}
  I(\bQ) = \left|
\sum_{p\kappa} 
f_\kappa(\bQ) 
e^{\iu\bQ\cdot
\left[\bR_p+\boldsymbol{\tau}_\kappa+\Delta\boldsymbol{\tau}_{p\kappa}\right]}
\right|^2 
\end{align}
where $\boldsymbol{\tau}_\kappa$ denotes the equilibrium coordinates of 
the nuclei, $ f_\kappa $ the atomic form factor of atom $ \kappa $,
and $\bQ$ the scattering wave vector. 
$ \Delta\boldsymbol{\tau}_{p\kappa}$ is the displacement of 
the atoms from their equilibrium configuration at the instant of the 
scattering process. 

At zero temperature and in absence of polarons, 
the atomic coordinates coincide with the equilibrium ones  
($\Delta\boldsymbol{\tau}_{p\kappa} =0$). 
This condition substantially simplifies the 
expression for the scattered intensity, leading to 
$\sum_{p\kappa} f_\kappa(\bQ)
e^{\iu\bQ\left[\bR_p+\boldsymbol{\tau}_\kappa\right]} 
\propto \sum_\kappa f_\kappa(\bQ) \delta_{\bQ, {\bf G}}$. 
Here, the function $\delta_{\bQ, {\bf G}}$ is 1 if the momentum transfer $\bQ$ coincides
with a reciprocal lattice vector $ {\bf G}$, and zero otherwise. 
In this limit, the scattered intensity is non-zero only for 
transferred momenta coinciding with the reciprocal lattice vectors, 
which reflects the suppression of inelastic scattering processes 
at zero temperature.  

At finite temperature, to first order in phonon scattering and in the basis of phonon normal modes \cite{Britt2023}, 
we can obtain the \emph{single-phonon diffuse scattering} terms:
\begin{equation}
  \label{eqn:eq_scat}
  I_1(\mathbf{Q}) \equiv \sum_\nu I^\nu_1(\bQ) \propto\sum_{\nu}\frac{n_{\mathbf{q}\nu}+1/2}{\omega_{\mathbf{q}\nu}}
    \big|\mathfrak{F}_{1\nu}(\mathbf{Q})\big|^2 
\end{equation}
where $\bf{q}$ is the reduced phonon wavevector (i.e. $\bf{q}$ = $\bf{Q}$ - $\bf{H}$, where $\bf{H}$ is the closest Bragg peak), and $n_{\bq\nu}$ is the mode-resolved phonon occupancy (BE distributed in thermal equilibrium). As this single phonon-diffuse scattering term depends solely on phonons at wavevector $\bq$, it is a direct measure of the phonon occupancy distribution at that wavevector, see \Cref{fig:schematic}. 

The $\mathfrak{F}_{1\nu}$ are known as the \emph{single-phonon structure factors}, geometric weights that describe the relative strength of scattering from different phonon modes:
\begin{equation}
    \label{eqn:f1nu}
    |\mathfrak{F}_{1\nu}(\mathbf{Q})|^2 = \bigg|\sum_\kappa  e^{-W_\kappa (\mathbf{Q})}\frac{f_\kappa (\mathbf{Q})}{\sqrt{M_\kappa }}(\mathbf{Q}\cdot \mathbf{e}_{\bq\nu\kappa })\bigg|^2
\end{equation}
where $W_\kappa$ is the exponent of the Debye-Waller (DW) factor \cite{Debye1913,Waller1923}. The critical term here is the dot product $(\mathbf{Q}\cdot \mathbf{e}_{\bq\nu\kappa })$, which shows that the contribution of a particular phonon mode is enhanced (suppressed) when the corresponding atomic polarization(s) are parallel (perpendicular) to the scattering vector, showing a sensitive dependence on the atomic polarization vectors $\mathbf{e}_{\bq\nu\kappa}$\cite{RenedeCotret2019}. 
\subsection{Diffuse Scattering of Polarons}
To account for the influence of a static polaronic distortion on the diffuse
scattering intensity at zero temperature, we identify the nuclear displacement
$\Delta {\boldsymbol \tau}_{\kappa p}$ in Eq.~\eqref{eqn:I} with the structural
changes induced by the polaron formation $\Delta {\boldsymbol \tau}_{\kappa
p}^{\rm pol}$ in Eq.~\eqref{eq:pol2}, with envelop functions $B_{\mathbf{q}
\nu}$ obtained from the solution of the self-consistent polaron equations
[Eq.~\eqref{eqn:polaron}]. 
Correspondingly, we express the changes of scattering intensity as: 
\begin{align}  
  I_\mathrm{pol}(\bQ) = \left|
\sum_{p\kappa}
f_\kappa(\bQ)
e^{\iu\bQ\cdot
\left[\bR_p+\boldsymbol{\tau}_\kappa+\Delta\boldsymbol{\tau}^{\rm pol}_{p\kappa}\right]}
\right|^2
\label{eqn:I_POL}
\end{align}
This definition enables us to promptly integrate polaronic effects in
\emph{ab-initio} simulations of inelastic scattering experiments. 
 
The extension of this approach to finite temperature 
can be conducted in close analogy to Ref.~\cite{Zacharias2021_PRB} 
based on the special displacement method. 
We evaluate the finite temperature inelastic scattering intensity in the absence of a polaron as:
\begin{align}
  I_\mathrm{ZG}(\bQ;T) = \left| \sum_{p \kappa}f_k(\bQ)e^{\iu\bQ\cdot\left[\bR_p+\boldsymbol{\tau}_\kappa+\Delta\boldsymbol{\tau}^\mathrm{ZG}_{p \kappa}\right]} \right|^2
  \label{eqn:I_ZG}
\end{align}
Here, the quantity $\Delta\boldsymbol{\tau}^\mathrm{ZG}_{p\kappa}$ 
denotes a single configuration of atomic displacements in a supercell 
which is determiend according to special displacement method by 
Zacharias and Giustino according to the procedure outlined in Ref.~\cite{Zacharias2021_PRB}. 
This approach enables to sample thermal effects 
via a single configuration. In the limit of large supercells, 
this approach it is equivalent to the William-Lax thermal average \cite{Zacharias2020} of the scattering 
intensity. We evaluate the finite temperature inelastic scattering intensity in presence of a polaron as: 
\begin{align}
  I_\mathrm{ZG+pol}(\bQ;T) = \left| \sum_{p \kappa}f_k(\bQ)e^{\iu\bQ\cdot\left[\bR_p+\boldsymbol{\tau}_\kappa+\Delta\boldsymbol{\tau}^\mathrm{ZG}_{p \kappa}+\Delta\boldsymbol{\tau}^\mathrm{pol}_{p \kappa}\right]} \right|^2
  \label{eqn:I_ZG_POL}
\end{align}
In this approach, finite temperature effects of the renormalization of 
polaron envelop functions are neglected. 

In this work, we numerically solve \Cref{eqn:eq_scat,eqn:I_POL,eqn:I_ZG_POL} 
to determine the diffuse scattering signatures of polarons for the limiting cases 
of small and large polarons. 
Our numerical investigation focuses on LiF, which is known to host
both electron and hole polarons in equilbrium \cite{Sadigh2015,Karsai2014,Pederson1988,Mallia2001,Shluger1993,Gavartin2003,Schirmer2006, Miller2003, Sio2019_PRL}.

\begin{figure*}[!t]
  \includegraphics[width=\linewidth]{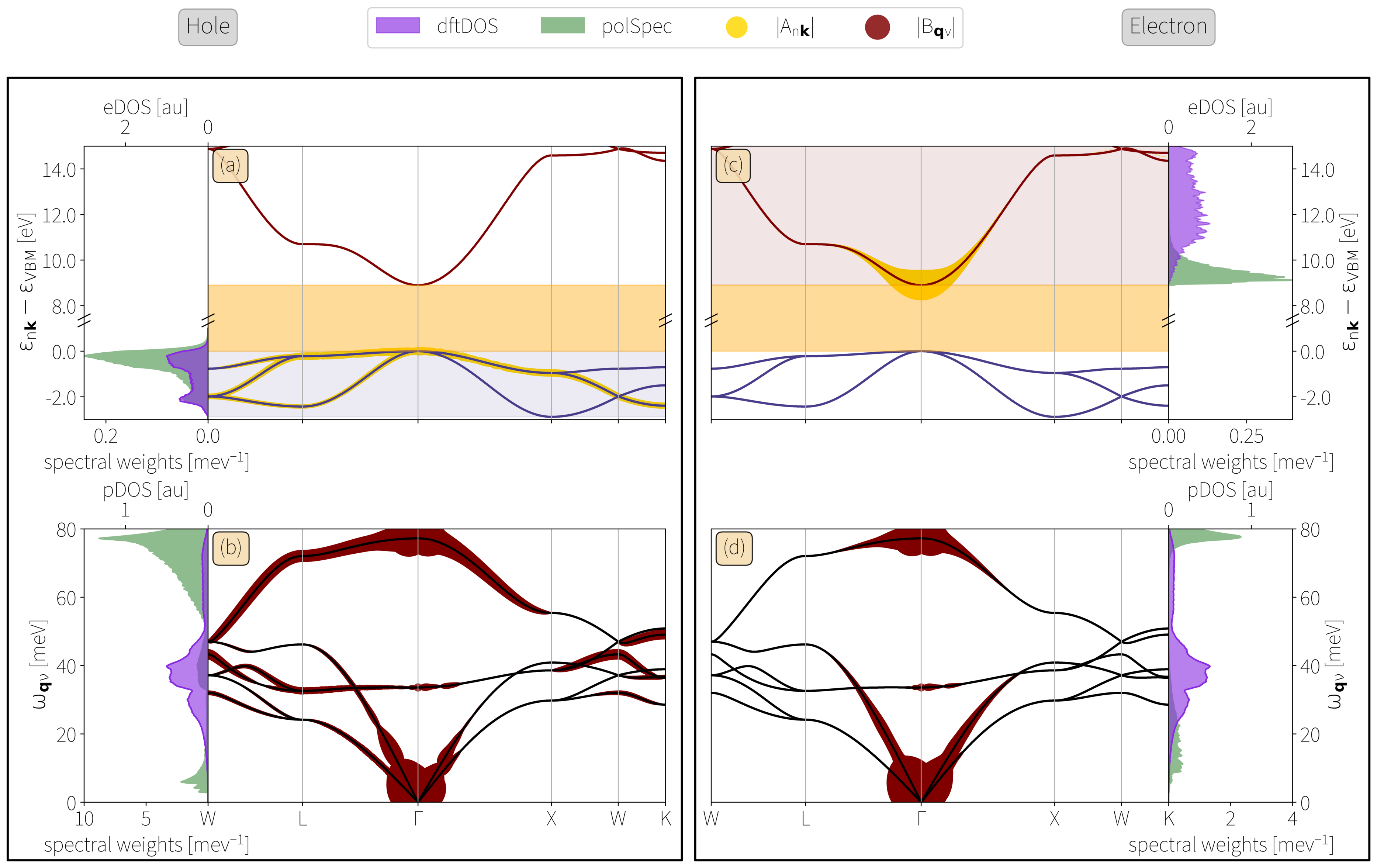}
  \caption{Electronic and vibrational momentum-resolved weights needed for polaron formation according to \Cref{eqn:polaron} in the insulator LiF. The electronic band structure and phonon dispersions are shown. The thickness of the bands at that particular point in the band is directionally proportional to the magnitude of the weights. The electronic weights, $A_{n\bk}$, are given by the yellow coloration on the bands, and the vibrational weights, $B_{\bq\nu}$, are given by the dark red coloration on the phonon dispersions. The weights are given for the electron system and phonon system for the hole polaron in (a) and (b), and for the electron polaron in (c) and (d). For the electronic weights, the bands included in the self-consistent calculation are those encompassed within the appropriate shaded region (dark blue for valence states in the hole polaron, maroon for conduction states in the electron polaron). The electronic bands have the band-gap region shaded in orange. Next to each primary axis is the DFT density of states (purple, arbitrary units) with the polaron spectral weight for the electronic and phononic system (green, meV$^{-1}$).}
  \label{fig:LiF_polaron_weights}
\end{figure*}
\subsection{Relation to the polaron envelope function}
In the following, we demonstrate that the change of
diffuse scattering intensity introduced by the formation
of polarons admits a simple representation in terms
of the polaron envelop function $B_{\bq\nu}$.
A direct consequence of this result is that
inelastic scattering constitutes a direct
route to directly probe polarons in crystals, and
it should in principle be capable of providing information on the
degree of localization of polaronic distortions (polaron radius).

To illustrate these points, we focus for simplicity on the low-temperature limit ($T=0$)
and we begin by rewriting the scattering intensity  in presence of a polaron by
a Taylor expansion of the exponential up to second order:
\begin{widetext}

\begin{align}   \label{eqn:I2}
  I_{\rm pol}(\bQ)
= &\sum_{\substack{pp^\prime\\\kappa \kappa^\prime}} 
f_\kappa(\bQ)
f_{\kappa^\prime}^*(\bQ)
e^{\iu\bQ\cdot
[(\bR_p - \bR_{p^\prime})
+(\boldsymbol{\tau}_\kappa - \boldsymbol{\tau}_{\kappa^\prime} ) ] } 
\big[1 + i \bQ \cdot(\Delta \boldsymbol{\tau}_{p\kappa}^{\rm pol}
-\Delta \boldsymbol{\tau}_{p^\prime\kappa^\prime}^{\rm pol} ) 
-\frac{1}{2}[ \bQ \cdot (\Delta \boldsymbol{\tau}_{p\kappa}^{\rm pol}
-\Delta \boldsymbol{\tau}_{p^\prime\kappa^\prime}^{\rm pol} ) ]^2\big]
\end{align}
The lowest-order term in the expansion coincides with  $I_0(\bQ)$,
the contribution of a static lattice to the scattering intensity
in the absence of polarons.
It can be easily verified that
the first-order term in the expansion only contributes to elastic scattering
processes, and can thus only lead to a renormalization of the
Debye-Waller factor.
The second-order term, conversely,  is the lowest-order contribution to the diffuse
scattering intensity.

As we are primarily interested in diffuse scattering, we omit the first-order term henceforth,
and retain only second-order terms in the following discussion.
Correspondingly, the change of diffuse scattering intensity due to the formation of polaron
can be expressed as:
\begin{align}   \label{eqn:DI}
\Delta  I_{\rm pol}(\bQ)
\equiv
 I_{\rm pol}(\bQ) -  I_{0}(\bQ)
= -\frac{1} {2}\sum_{\substack{pp^\prime\\\kappa \kappa^\prime}} 
f_\kappa(\bQ)
f_{\kappa'}^*(\bQ)
e^{\iu\bQ\cdot
[(\bR_p - \bR_{p'})
+(\boldsymbol{\tau}_\kappa - \boldsymbol{\tau}_{\kappa'} ) ] }
\big[ \bQ \cdot (\Delta \boldsymbol{\tau}_{\kappa p}^{\rm pol}
-\Delta \boldsymbol{\tau}_{\kappa'p'} ^{\rm pol}) \big]^2
\end{align}
After a few algebraic manipulation, we deduce
an explicit expression of $\Delta  I_{\rm pol}(\bQ)$ in terms of the polaron envelop function:
\begin{align}
\Delta I _{\rm pol}(\bQ) &=
- \hbar{\rm Re}   \sum_{\nu\nu'}
\frac{ B_{\bq \nu} B_{\mathbf{q} \nu'}^*}
{\sqrt { \omega_{\bq \nu}\omega_{\bq \nu'}} }
  \sum_{\kappa \kappa'}
f_\kappa(\bQ)
f_{\kappa'}^*(\bQ)
e^{\iu\bQ\cdot
[
(\boldsymbol{\tau}_\kappa - \boldsymbol{\tau}_{\kappa'} ) ] }
\frac{( \bQ \cdot {\bf e}_{ \mathbf{q} \nu^\prime\kappa^\prime} )( \bQ \cdot {\bf e}^{*}_{ \bq \nu\kappa} ) }
{ \sqrt{ M_\kappa M_{\kappa^\prime} } }
\end{align}
\end{widetext}
This result has been obtained by combining Eqs.~\eqref{eqn:DI} and \eqref{eq:pol2} with
the Born-von-Kármán sum rule ($\sum_p e^{i \bq \bR_p} = N_p \delta_{\bq,0}^{\bf G}$), and
making use of the conditions $ B_{-\bq \nu} =  [B_{\bq \nu}]^*$ and ${\bf e}_{ \mathbf{-q} \nu\kappa} =[ {\bf e}_{ \mathbf{q} \nu\kappa}]^*$.

For materials in which polarons are associated with structural distortions along a single
polar phonon -- as, e.g., in the case of LiF --  this expression simplifies further,
and we can express the change of diffuse scattering intensity induced by the polaron formation as:
\begin{align} \label{eq:Ipol_2nd}
\Delta I _{\rm pol}(\bQ) &=
-\frac{ | B_{\bq }|^2 }
{{ \omega_{\bq }} } |\mathbb{F}(\bQ)|^2
\end{align}
where we introduced the function:
\begin{align}
|\mathbb{F}(\bQ)|^2 =
\hbar \left |
\sum_\kappa
e^{\iu\bQ \cdot \boldsymbol{\tau}_\kappa}\frac{f_\kappa(\bQ)}{\sqrt{ M_\kappa}}  ( \bQ \cdot {\bf e}_{ \mathbf{q}\kappa } )
\right|^2
\end{align}
analogously to the single-phonon structure factor in Eq.~\eqref{eqn:f1nu}. Overall, Eq.~\eqref{eq:Ipol_2nd} indicates the changes of diffuse scattering intensity
induced by the formation of a polaron are directly proportional to the polaron envelop function.
This simple result establishes a direct link between polaronic distortion in crystals
and their fingerprints in diffuse scattering experiments.

\begin{figure*}[!th]
  \includegraphics[width=\linewidth]{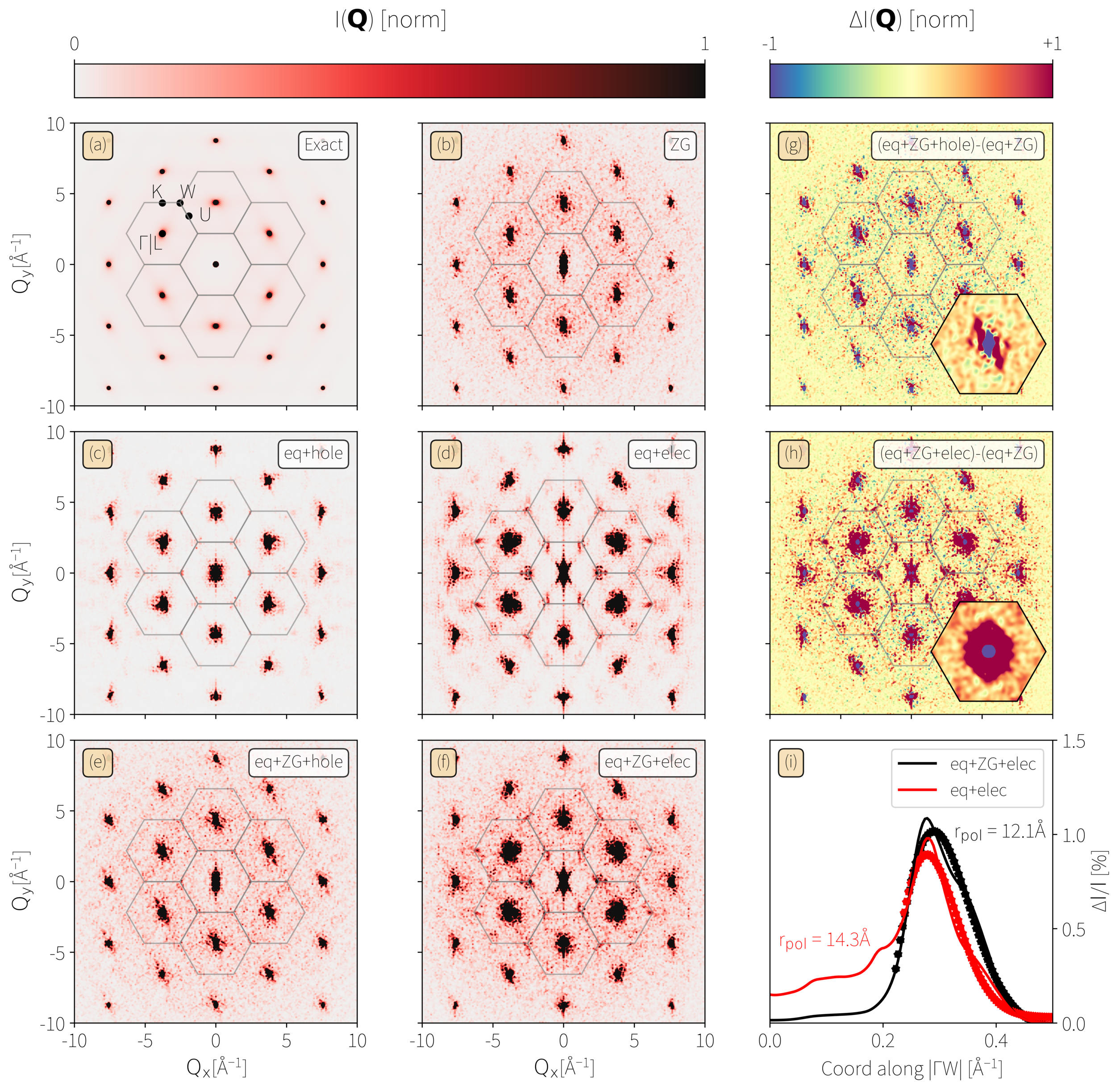}
  \caption{Polaron-diffuse scattering in \ce{LiF}. All diffuse scattering patterns are for scattering in the $[111]$ direction of the FCC lattice, thus yielding six-fold symmetry. The exact diffuse scattering intensity from \Cref{eqn:eq_scat} is given in panel (a) with the high symmetry points of the BZ from the $[111]$ projection labelled, and the ZG scattering intensity is in panel (b). The static polaron scattering intensity from \Cref{eqn:I_POL} is given for the hole polaron in panel (c) and for the electron polaron in panel (d), with the inclusion of thermal effects via the ZG displacements in \Cref{eqn:I_ZG_POL} in panels (e) and (f) respectively. The differential diffuse intensity, given by $\Delta I=I_\mathrm{ZG+pol}-I_\mathrm{ZG}$, for the hole and electron polarons are given in panels (g) and (h) respectively. Inset in these panels are the average of the first-order reflections across the entire BZ. Panel (i) gives the linecut of relative differential intensity for the electron polaron, with the radius of the polaron determined via the point defect model shown.}
  \label{fig:LiF_polaron_scattering}
 \end{figure*}

\section{Polarons in \ce{LiF}} \label{sec:LiF}  
We solve the self-consistent \Cref{eqn:polaron} and render the sets of weights $\{A_{n\bk}\}$ and $\{B_{\bq\nu}\}$ in \Cref{fig:LiF_polaron_weights}. Both such polarons are found to exist with formation energies and real-space extents matching those reported in Ref.~\cite{Sio2019}. Namely, the momentum-resolved weights of the hole polaron are dispersed throughout the BZ, consistent with a small real-space polaron with extracted real-space spread of $2r_\mathrm{pol}^{\mathrm{hole}}=$ \qty{0.93}{\AA}, and using Makov-Payne extrapolation \cite{Makov1995}, we find a formation energy of \qty{-1.978}{eV}. Conversely, the electron polaron has weights highly localized in the BZ, consistent with a large real-space polaron with an extracted real-space spread of $2r_\mathrm{pol}^\mathrm{elec}=$ \qty{8.59}{\AA}, and an extrapolated formation energy of \qty{-268}{meV}. See SI for further details.

\subsection{Polaron-Diffuse Scattering in \ce{LiF}}
We thus proceed to determine the expected polaron-diffuse scattering signatures as measurable by UED and UEDS. Computing the hole-polaron-diffuse scattering pattern via \Cref{eqn:I_ZG_POL}, as in \Cref{fig:LiF_polaron_scattering}c and e, shows a small anisotropic change compared to equilibrium scattering, shown in \Cref{fig:LiF_polaron_scattering}b. The low intensity of the differential scattering in \Cref{fig:LiF_polaron_scattering}g owes to the small spatial extent of the polaron, and the anisotropy with respect to phonon momentum results from the anisotropic $p$-like orbital character of the hole polaron atomic displacements and the corresponding wavefunction (see SI) that extend along the diagonal of the projected [111] scattering. However, the electron polaron (shown in \Cref{fig:LiF_polaron_scattering}d and f), extending over many unit cells, shows a differential increase in diffuse intensity in a ring about the Bragg peaks (\Cref{fig:LiF_polaron_scattering}h). 

The point defect model \cite{RenedeCotret2022} shows that an atomic displacement field with Gaussian distributed values about some charge trapping center ($\bu(\br)\propto e^{|\br|/r_\mathrm{pol}^2}\hat{\bu}$) would result in an annular momentum dependence to the diffuse scattering about a Bragg peak. Namely:
\begin{equation}
    \frac{I(\Delta\boldsymbol{\tau}_{p\kappa}^\mathrm{pol})-I(\Delta\boldsymbol{\tau}_{p\kappa}=0)}{I(\Delta\boldsymbol{\tau}_{p\kappa}=0)}\propto |\bq|r^2_\mathrm{pol} e^{-\nicefrac{|\bq|^2r^2_\mathrm{pol}}{2}}
\end{equation}
In this instance, the electron-polaron-diffuse scattering qualitatively matches the point defect model indicated by the Gaussian-like distribution of atomic displacements (see SI) about the charge trapping center. We show the linecuts along the $|\Gamma W|$ path for the electron polaron in both $I_\mathrm{ZG+pol}$ and $I_\mathrm{pol}$ in \Cref{fig:LiF_polaron_scattering}g, where the fit to the point defect model extracts values of \qty{12-14}{\AA}, matching closely the true value of \qty{8.56}{\AA}. 

\begin{figure}[!t]
    \centering
    \includegraphics[width=\linewidth]{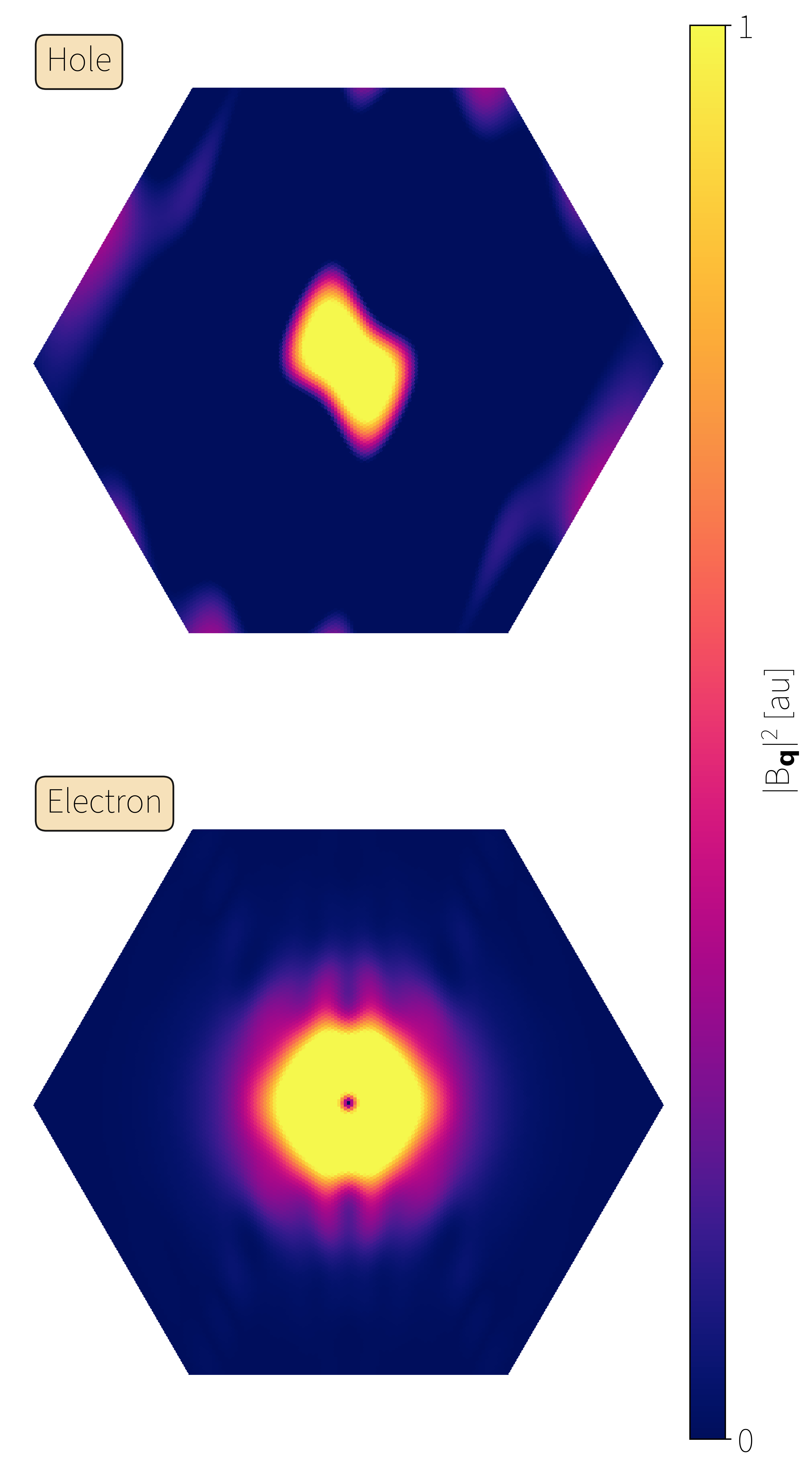}
    \caption{The contribution to the polaron envelope function from the polar optical mode in \ce{LiF} for the hole (top) and electron (bottom) polaron. }
    \label{fig:Bqv}
\end{figure}
Further, as there is only a single polar phonon primarily responsible for polaron formation in this material, there will be, according to \Cref{eq:Ipol_2nd}, a direct connection between the differential diffuse scattering of the polaron and the contribution of this mode $|B_{\bq}|^2$ to the polaron envelope function. We visualize the momentum-resolved contribution of this polar mode to the hole and electron polarons in \Cref{fig:Bqv}, where the momentum dependence of $|B_{\bq}|^2$ and that of the differential polaron scattering $\Delta I_\mathrm{pol}$ in \Cref{fig:LiF_polaron_scattering}g-h match identically.

The polaronic signatures of UEDS so far have relied on charge carriers independently coupling to the lattice to create a localised ionic distortion. Yet it is possible, for materials that have a large interaction strength between bound electron-hole pairs and phonons, that the ions of a material can be distorted to form a self-trapped exciton \cite{Williams1990}, or \enquote{exciton-polaron}. To describe the exciton-polaron necessitates moving beyond the single-particle picture to a fully many-body theory that is currently the subject of intense research \cite{Zhenbang2024}, but is beyond the scope of this work.

\section{Conclusion} \label{sec:con}
We have show that a description of polaron formation within the basis of Kohn-Sham orbitals and phonon normal modes allows for the rapid identification of diffuse scattering signatures of polarons within Laval-Born-James theory and the relation of experimentally accessible diffuse scattering images directly to the underlying envelope of the polaron wavefunction. We have extended this formalism to include finite-temperature phonon-assisted scattering, and validate the approach on a prototypical ionic insulator \ce{LiF}. Here, we discover that the electron polaron, with large spatial extent, can be reasonably described as a point defect in the material, and admits an annular momentum dependence to the resulting scattering about zone center. Both polaron scattering images are shown to directly reveal the underlying polaron wavefunction with full momentum dependence. As UEDS is already able to detect the dynamics of EPC-mediated scattering processes, now being able probe the structure of polarons via their scattering signatures allows for the possiblility of time-resolving the dynamics of polaron formation, and understanding further the origins of these complicated objects in real materials.

\section*{Methods}\label{sec:methods}
The calculations used in this work relied on the computational suite \texttt{QuantumESPRESSO} \cite{Giannozzi2009, Giannozzi2017}, using norm-conserving Troullier-Martins pseudopotentials \cite{Troullier1991} and the Perdew-Burke-Ernzerhof generalized gradient approximation for the exchange-correlation functional \cite{Perdew1996}. We used an energy cutoff of \qty{150}{Ry}, and a $12\times 12\times 12$ Monkhorst-Pack $\bk$-grid for the electronic structure calculations. Second-order force constants were computed using density functional perturbation theory on a $12\times 12\times 12$ $\bq$-grid.

The determination of the polaron weights, as well as the EPC calculations and the polaron diffuse scattering patterns, were computed using developer versions of the \texttt{EPW} suite \cite{Giustino2007, Ponce2016, Lee2023} using the maximally-localized Wannier functions (MLWF) computed by \texttt{wannier90} \cite{Pizzi2020} and the self-consistent polaron equations of Sio \emph{et al} \cite{Sio2019}. Polar corrections to the MLWF interpolation of the EPC matrix elements were explicitly implemented to account for short- and long-range interactions \cite{Verdi2015}. The EPC matrix elements were explicitly computed using the original $12\times 12\times 12$ $\bk$-grid and $12\times 12\times 12$ $\bq$-grid. 

The polaron weights $A_{n\bk}$ and $B_{\bq\nu}$ were explicitly solved on $24\times 24\times 24$ $\bk$- and $\bq$-grids by instantiating the electron weights $A^0_{n\bk}$ with Gaussian distribution with an \emph{a-priori} reasonable polaron size about the valence band maximum (conduction band minimum) for the hole (electron) polaron. \Cref{eqn:polaron} were solve self-consistently until the magnitude of the difference in ionic displacements across the supercell between iterations $n-1$ and $n$ was below a set tolerance, namely $\max_{p\kappa}|\Delta\boldsymbol{\tau}_{p\kappa}^{n}-\Delta\boldsymbol{\tau}_{p\kappa}^{n-1}|<\delta\tau$. The center position of the starting wavepacket $A^0_{n\bk}$ was iterated across the BZ to ensure global convergence. The Makov-Payne extrapolation of polaron formation energies was performed by calculating the formation energy of the hole polaron on $12\times 12\times 12$, $24\times 24\times 24$, and $32\times 32\times 32$ supercells, and $18\times 18\times 18$, $24\times 24\times 24$, and $32\times 32\times 32$ supercells for the electron polaron. All diffuse scattering calculations utilized developer versions of the \texttt{disca.x} and \texttt{ZG.x} codes in the \texttt{EPW} suite  \cite{Zacharias2021_PRB}.
\begin{acknowledgments}
  This work was supported by the Natural Sciences and Engineering Research Council of Canada (NSERC), the Fonds de Recherche du Québec-Nature et Technologies (FRQNT), the Canada Foundation for Innovation (CFI), Quantum Photonics Quebec (PQ2), and a McGill Fessenden Professorship. 
  F.C. acknowledges funding from the Deutsche Forschungsgemeinschaft (DFG), Projects No. 499426961 and No. 434434223. 
  B.J.S. conceived the research.  
\end{acknowledgments}
\bibliography{references}

\end{document}